\documentclass[fleqn,12pt,twoside]{article}
\usepackage{espcrc1}

\usepackage{graphicx}
\usepackage[figuresright]{rotating}

\newcommand{\AmS}{{\protect\the\textfont2
  A\kern-.1667em\lower.5ex\hbox{M}\kern-.125emS}}

\title{Strange Fluctuations at RHIC}

\author{M. Abdel-Aziz and S. Gavin\address[WSU]{Physics Department, Wayne State
        University, Detroit, MI 48201}}

\begin{document}

\maketitle

\begin{abstract}
Robust statistical observables can be used to extract the novel
isospin fluctuations from background contributions in $K_S^0K^\pm$
measurements in nuclear collisions. To illustrate how this can be
done, we present new HIJING and UrQMD computations of these
observables.
\end{abstract}

\section{INTRODUCTION}

Relativistic nuclear collisions can produce matter in which chiral
symmetry is restored. One possible consequence of this restoration
is the formation of disoriented chiral condensates (DCC) --
transient regions in which the average chiral order parameter
differs from its value in the surrounding vacuum. Previous efforts
to describe DCC signals have focused on pion production. In
Ref.~\cite{GavinKapusta}, Gavin and Kapusta explored the possible
influence of DCC on kaon production, inspired by an explanation by
Kapusta and Wong \cite{KapustaWong} of measurements of $\Omega$
and $\overline\Omega$ baryon enhancement at the CERN SPS, as
discussed by Wong in these proceedings.

If true, Ref.~\cite{KapustaWong} implies that the evolution of the
condensate has a substantial strange component and, therefore, can
produce kaon isospin fluctuations. A further implication is that
the DCC regions must be rather small, with a size of about $2$~fm,
in accord with predictions based on dynamical simulations of the
two flavor linear sigma model. Such a size is consistent with WA98
results reported here, which reveal no evidence of large DCC
domains in neutral and charged pions. In this talk we study kaon
isospin fluctuations in the presence of many small DCC; both kaon
and pion fluctuations were addressed in \cite{GavinKapusta}. In
the next section we discuss how DCC may lead to kaon fluctuations
and propose statistical observables to detect them. In sec.~3 we
present work in progress on event generator simulations to
understand the magnitude and centrality dependence of these
observables in the absence of DCC.

\section{STRANGE DCC}

To illustrate how a strange DCC can form, we first consider QCD
with only up and down quark flavors. Equilibrium high temperature
QCD respects chiral symmetry if the quarks are taken to be
massless. This symmetry is broken below $T_c\sim 150$~MeV by the
formation of a chiral condensate $\sigma$ that is a scalar isopin
singlet. However, chiral symmetry implies that $\sigma$ is
degenerate with a pseudoscalar isospin triplet of fields with the
same quantum numbers as the pions.

DCC can form when a heavy ion collision produces a high energy
density quark-gluon system that then rapidly expands and cools
through the critical temperature. Such a system can initially
break chiral symmetry along one of the pion directions, but must
then evolve to the $T=0$ vacuum by radiating pions. A single
coherent DCC radiates a fraction $f_\pi$ of neutral pions that
satisfies the probability distribution $\rho_1(f_\pi) =
\{2f_\pi^{1/2}\}^{-1}$. Such isospin fluctuations constitute the
primary signal for DCC formation in the pion sector. The
enhancement of baryon-antibaryon pair production in
\cite{KapustaWong} is a secondary effect due to the relation
between baryon number and the topology of the pion condensate
field.

This two flavor idealization only applies if the strange quark
mass $m_s$ can be taken to be infinite. Alternatively, if we take
$m_s = m_u = m_d =0$, then the chiral condensate would be an
up-down-strange symmetric scalar field. For a realistic $m_s$, a
disoriented condensate can evolve by radiating $\pi, K, \eta$ and
$\eta^\prime$ mesons. Kaon fluctuations from a single large DCC
would satisfy $\rho_1(f_K) = 1$, where $f_K = (K^0 +
\overline{K}^0)/(K^+ + K^- + K^0 + \overline{K}^0)$
\cite{Randrup}. Moreover, the condensate fluctuations can now
produce strange baryon pairs \cite{KapustaWong}. Linear sigma
model simulations indicate that pion fluctuations dominate
three-flavor DCC behavior, while the fraction of energy imparted
to kaon fluctuations is very small due to the kaons' larger mass.
On the other hand, domain formation may be induced by other
mechanisms such as 
bubble formation or decay of the Polyakov loop condensate
\cite{Pisarski}.

Detection of small incoherent DCC regions in high energy heavy ion
collisions requires a statistical analysis in the $\pi^0\pi^\pm$
or the $K_S^0K^\pm$ channels. Neutral mesons can be detected by
the decays $\pi^0\rightarrow \gamma\gamma$ or $K^0_S\rightarrow
\pi^+\pi^-$. The analysis we propose in \cite{GavinKapusta} is
sensitive to correlations due to isospin fluctuations. We expect
these correlations to vary when DCC regions increase in abundance
or size as centrality. Correlations of $K_S^0 K^\pm$ are reflected
in the isospin variance,
\begin{equation}
\nu = \left\langle\left({{N_0}\over{\langle N_0\rangle}} -
{{N_c}\over{\langle N_c\rangle}} \right)^2\right\rangle,
\end{equation}
where $N_0$ and $N_c$ are the number of neutral and charged
mesons. For a single DCC, we find $\nu \rightarrow 4\langle
(2f_K-1)^2\rangle = 4/3$. In contrast, an uncorrelated thermal
source yields
\begin{equation}
\nu_{\rm stat} = \langle N_0\rangle^{-1} + \langle N_c\rangle^{-1}
= 4\langle N_K\rangle^{-1},
\end{equation}
with $N_K = N_c + N_0$. The dynamic isospin observable
\cite{GavinKapusta}
\begin{equation}
\nu_{\rm dyn} = \nu - \nu_{\rm stat}
\end{equation}
is convenient for experimenters, because this quantity is
independent of detection efficiency \cite{Pruneau} as discussed in
Pruneau's talk. Robust observables are useful for DCC studies
because charged and neutral particles are identified using very
different techniques and, consequently, are detected with
different efficiency.

Why does the DCC's size matter? Searches for structure in
individual events -- as conducted by WA98 -- can distinguish DCC
isospin fluctuations from a thermal background only if the
disoriented region is sufficiently large. DCC can then be the
dominant source of pions at low transverse momenta, since $\langle
p_t \rangle\sim 1/R$ for a coherent region of size $R$.
Experiments focusing on low $p_t$ can then find anomalous events.
For small domains as inferred in \cite{KapustaWong}, this event
structure is hard to distinguish from conventional fluctuations.
However, statistical observables such as (3) can signal even a
small DCC contribution for a sufficiently large event sample.

In Ref.~\cite{GavinKapusta}, we compute the exact distribution of
kaons due to many small DCC regions. For $n \gg 1$ domains, this
distribution tends toward a Gaussian of mean $\langle f\rangle
=1/2$ and isospin variance $\nu_{\rm dcc} = 4/3n$. In a nuclear
collision, the best we can hope is that a fraction $\beta$ of
kaons come from the decay or realignment of DCC domains, with the
remainder $\alpha = 1-\beta$ from thermal sources. Folding
together the thermal and the $n$-DCC distributions, we find $\nu =
\alpha^2\nu_{\rm dcc}+\beta^2\nu_{\rm stat}$, which implies
\begin{equation}
\nu_{\rm dyn} = 4\beta(\beta/3n - 1/\langle N_K\rangle).
\end{equation}
This quantity can be positive or negative depending on the
magnitude of $\beta$ compared to the number of domains per kaon.

\section{FLUCTUATIONS IN COLLISIONS}

We now discuss work in progress using event generators to simulate
conventional sources of kaon fluctuations. In the absence of DCC,
$\beta = 0$ so that $\nu_{\rm dyn}^{c0} \equiv 0$, see eq.~(4). On
the other hand, incomplete equilibration in nuclear collisions can
result in dynamical correlations not described above. In
particular, correlations at the $NN$ level can affect $\nu_{\rm
dyn}$. To explore how this might occur, we use HIJING and UrQMD to
estimate the influence of conventional collision geometry and
dynamics on the centrality dependence.
\begin{figure}
\centerline{\includegraphics[width=5in]{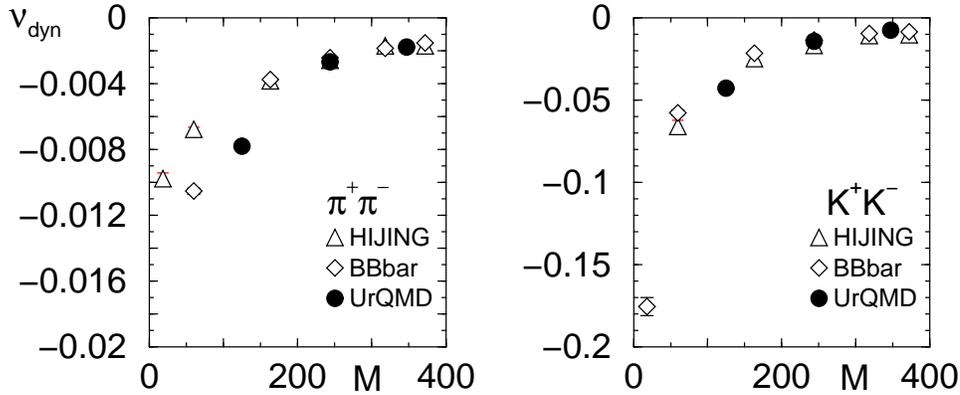}} 
\vskip -0.4in
\caption[]{Dynamic charge fluctuations for $\pi^+\pi^-$ (left) and
$K^+K^-$ (right).}\end{figure}
To establish the relevance of these models, we show in fig.~1 the
dynamic \emph{charge} fluctuations $\nu_{\rm dyn}^{+-}$ for pions
and kaons as functions of the number of participants $M$ computed
from 300,000 HIJING, HIJING/$B\overline{B}$ and 50,000 UrQMD
events for Au+Au at 200 A$GeV$ in the rapidity range $-0.5<y<0.5$.
The uncertainties shown are statistical. Observe that all models
give nearly the same values of $\nu_{\rm dyn}^{+-}$. STAR data in
Pruneau's talk in these proceedings is in reasonable agreement
with this data.

Figure 2 (left) shows the dynamic isospin observable as a function
of $M$ from HIJING, HIJING/$B\overline{B}$ and UrQMD. The large
disagreement between the HIJING variants and UrQMD is astonishing,
given the similarity of the charge fluctuations. To understand the
contribution of collision dynamics to $\nu_{\rm dyn}^{c0}$, we
compare our simulations to the wounded nucleon model (WNM). This
model implies $\nu_{\rm wnm} \approx 2\nu_{pp}/M$, where $M$ is
the number of participants. The value in pp collisions $\nu_{pp}$
is determined from pp simulations. The WNM curves in fig.~2 agree
with HIJING and HIJING/$B\overline{B}$ (the upper sets of points),
but disagree with central UrQMD (the lower points). These trends
may stem from the inclusion of rescattering in UrQMD but not in
HIJING.
\begin{figure}
\centerline{\includegraphics[width=5in]{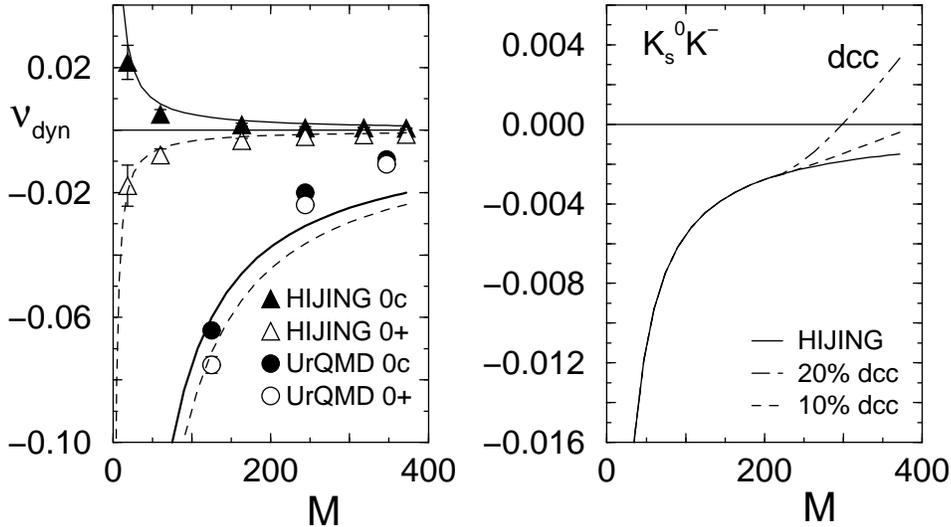}}
\vskip -0.4 in 
\caption[]{(left) Dynamic isospin fluctuations of
$K^c K_s^0$ and $K^+K_s^0$ are respectively compared to solid and
dashed curves computed using a wounded nucleon model normalized to
pp simulations ($K^c= K^+ + K^-$). (right) Effect of DCC.
}\end{figure}

To illustrate the possible scenario for the onset of DCC effects,
we estimate $\nu_{\rm dyn}$ by adding DCC and wounded-nucleon
contributions to the kaon variance and using (3) to find $\nu_{\rm
dyn} = \beta^2\nu_{\rm dcc} + (1-\beta)^2\nu_{\rm wnm}$. We assume
that the fraction of DCC kaons $\beta$ exhibits a threshold
behavior above an impact parameter $b_0$, $\beta =
\beta_0[1-(b/b_0)^2]$, where $b_0$ and $\beta_0$ are ad hoc
constants. In fig.~2, we show estimates assuming that 10 domains
contribute kaons in the range $-0.5 < y < 0.5$ for $b_0 \sim
6$~fm, taking the dcc fraction $\beta_0\sim 20\%$.

In summary, we have argued that measurements of $K_s^0K^\pm$
correlations may probe a variety of interesting phenomena,
especially 2+1 flavor DCC. The robust statistical variable
$\nu_{dyn}$ is sensitive to DCC even if domains are small. In
addition, we have estimated $\nu_{dyn}$ in the absence of DCC
using HIJING and UrQMD event generators. We find that isospin
fluctuations distinguish between these models, while charge
fluctuations do not.

We thank J. Kapusta for collaboration on \cite{GavinKapusta}, and
R. Bellwied, C. Pruneau and S. Voloshin for discussions. This work
is supported in part by the U.S. DOE grant DE-FG02-92ER40713.


\begin{thebibliography}{9}


\bibitem{GavinKapusta} S.~Gavin and J.~I.~Kapusta, Phys Rev C65, 054910 (2002);
nucl-th/0112083.

\bibitem{KapustaWong}
J.~I.~Kapusta and S.~M.~Wong,
Phys.\ Rev.\ Lett.\  {\bf 86}, 4251 (2001); nucl-th/0012006.

\bibitem{Pisarski} A. Dumitru and R. Pisarski, Phys.Lett. B504 (2001) 282; O.
Scavenius, A. Dumitru, and  J.T. Lenaghan, hep-ph/0201079

\bibitem{Pruneau} C. Pruneau, S. Gavin and S. Voloshin, Phys.\ Rev.\ C, to be published; nucl-ex/0204011.

\bibitem{Randrup} J.~Schaffner-Bielich and J.~Randrup, Phys.\ Rev.\ C
{\bf 59}, 3329 (1999); nucl-th/9812032.



\end{thebibliography}
\end{document}